\documentclass[useAMS]{mn2e}
\usepackage{graphicx}

\title[Cepheid distance to NGC 6822]{The Cepheid distance to the Local Group Galaxy NGC 6822}

\author[Feast et al.]{ Michael W. Feast$^{1,2}$, Patricia A. Whitelock$^{2,1}$,
John W. Menzies$^{2}$ and Noriyuki \newauthor Matsunaga$^3$\\
      $^1$ Astronomy, Cosmology and Gravity Centre, Astronomy Department,
           University of Cape Town, 7701 Rondebosch, South Africa.\\
      $^2$ South African Astronomical Observatory, P.O.Box 9, 7935
           Observatory, South Africa.\\
      $^3$ Kiso Observatory, Institute of Astronomy, The University of Tokyo,
           10762-30, Mitake, Kiso, Nagano 397-0101, Japan.\\
 }

\begin{document}
\maketitle
\begin{abstract}
Recent estimates of the Cepheid distance modulus of NGC 6822 differ by 0.18 mag. To investigate this
we present new multi-epoch $JHK_{S}$ photometry of classical Cepheids in 
the central region of NGC 6822 and show that there is a zero-point
difference from earlier work. These data together with optical and mid-infrared observations from the
literature are used to derive estimates of the distance modulus of NGC 6822. A best value of 23.40 mag is adopted,
based on an LMC distance modulus of 18.50 mag.
The standard error of this quantity is $\sim0.05$ mag. We show that to derive consistent moduli
from Cepheid observations at different wavelengths, it is necessary that the fiducial LMC period-luminosity
relations at these wavelengths should refer to the same subsample of stars. Such a set is provided. A distance modulus
based on RR Lyrae variables agrees with the Cepheid result. 

\end{abstract}
\section{Introduction}
Local Group galaxies, besides being of interest in their own right, are important testbeds of Galactic and
extragalactic distance indicators. For much extragalactic work classical Cepheids are of prime importance.
It is somewhat disconcerting therefore that recent work\footnote{Early work on the Cepheids in NGC 6822 is summarized by Madore et al. (2009a)}
on Cepheids in NGC 6822, a Local Group dwarf
galaxy, has led to distance moduli which differ by up to 0.18 mag,
a 9\% range in distance. This is at a time when there is a general expectation
that Cepheid distances to extragalactic systems can be obtained with very high precision
(e.g. a distance with a 3\% uncertainty for M31 (Riess et al. 2011)). Uncertainties in the Cepheid
distance to NGC 6822 have implications for the use of Cepheids generally.

For NGC 6822 Pietrzy\'{n}ski et al. (2004) found $ \rm (m-M)_{o}=23.34 \pm 0.04$ (statistical) $\pm 0.05$(systematic) mag using  a period-magnitude-colour relation in $VI$. Combining
these results with new $JK$ observations, Gieren et al. (2006) found $23.312\pm 0.021$ mag.
On the other hand, Madore et al. (2009a) combined optical observations with mid-infrared data to derive
a modulus of $23.49 \pm 0.03$ mag. 
All these estimates are based on an assumed modulus of the LMC of 18.50 mag
and it was also assumed that any Cepheid metallicity corrections between the LMC and NGC 6822 were negligible. 
Madore et al. showed that it was difficult
to combine their mid-infrared data with the Gieren et al. $JK$ results, suggesting possible problems
with these or the mid-infrared data.
 In the present paper we present new multi-epoch $JHK_{S}$ photometry of Cepheids in the central regions
of NGC 6822 and compare this with earlier work. We then derive the
distance modulus of the galaxy by combining these data with optical and mid-infrared observations in a
variety of ways and discuss the discrepancies noted above. 

This work is part of a $JHK_{S}$ study of Local Group galaxies aimed primarily at AGB variables, but also
dealing with other types of objects, and the structure of colour-magnitude diagrams.

\section{Observations}
Our survey of NGC 6822 is confined to the optical bar which is aligned nearly N-S. We used the Japanese-South African IRSF telescope equipped with the SIRIUS camera, which permits simultaneous imaging in the $J, H$ and $K_S$ bands. We defined 3 overlapping fields, with field 1 centred at $\alpha$(2000.0) = $19^h44^m56^s$ and $\delta$(2000.0) =$-14^o48'06''$. Fields 2 and 3 are centred 6.7 arcmin N and S, respectively, from field 1. The three fields, approximately 7.8 arcmin square, were observed in $JHK_S$ at 19, 18 and 16 epochs, respectively, over a period of 3.5 years. Typically 30 dithered exposures of 30 s each were combined at each epoch, though occasionally, depending on sky brightness at $K_S$, exposure times were reduced to 20 s. 

Photometry was carried out with the DoPHOT program (Schechter et al. 1993) in ''fixed-position" mode. To allow for possible non-photometric nights, a set of bright reference stars was used to normalise the resultant magnitudes for the images in each band. The mean magnitudes of stars in common with 2MASS were used to put our photometry onto the 2MASS system. Table 1 shows the number of 2MASS stars of quality AAA used in each field, together with the standard deviations (s.d.)  of the comparisons with our magnitudes. The comparison stars cover the range, 12.5 -- 16.1 mag in $J$,
12.1 -- 15.1 mag in $H$, and 12.0 -- 14.7 mag in $K_S$. For each field, there is a different number of stars in each band following rejection of $>3\sigma$ outliers.

\begin{table}
\caption[]{Comparison with 2MASS}
\begin{center}
\begin{tabular}{cccc}
Band & Field & No. stars & \multicolumn{1}{c}{s.d.} \\
& & & \multicolumn{1}{c}{mag} \\
\hline
$J$ & 1 &  86 & 0.053 \\
  & 2 & 89 & 0.044 \\
  & 3 &  84 & 0.035 \\
$H$ & 1 & 67 & 0.061 \\
  & 2 & 72 & 0.048 \\
  & 3 & 75 & 0.032 \\
$K$ & 1 & 75 & 0.074 \\
  & 2 & 69 & 0.084 \\
  & 3 & 71 & 0.051 \\
\hline
\end{tabular}
\end{center}
\label{tab_zp} 
\end{table}

The standard deviations are consistent with expectations, being almost entirely attributable to the quoted 2MASS errors. The residuals were investigated for colour equation, but none was found over the $J-K_S $ range, 0.4 to 1.5 mag, of the 2MASS stars. The implication is that our zero points are accurate to 0.01 mag in all bands.
\section{Results}
\begin{figure}
\includegraphics[width=8.5cm,viewport=0 0 600 300,clip]{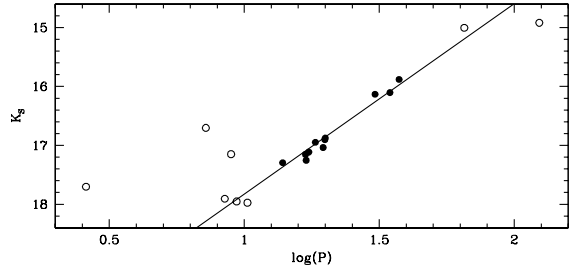}
\caption{Intensity mean $K_S$ magnitudes versus log Period. The line,
$K_S = -3.234log(P)+21.062$, is the best fit to the points for selected Cepheids (filled circles) as described in the text.
}
\label{fig_PL}
\end{figure}

Table \ref{tab_obs} lists our individual observations of
the Cepheids (or possible Cepheids) in common with Pietrzy\'{n}ski et al. (2004). Intensity mean magnitudes for these stars are listed in Table \ref{tab_mean}.
The mean magnitudes were determined by converting our $JHK_S$ magnitudes
to intensities then Fourier fitting a sine curve. The mean intensity from
the best fit was converted to a magnitude, which is listed in the table. Fourier
fits of up to third order (i.e. using as many as two harmonics) were
tried and the order that gave the best fit was used; for cep014 and
fainter stars only first order fits were made.
Fig. \ref{fig_PL} shows  our intensity mean values of $K_{S}$ plotted against log period.

\begin{table}
\caption{$JHK_S$ photometry for Cepheids observed in NGC 6822}
\begin{center}
\begin{tabular}{lllllll}
\hline
\multicolumn{1}{c}{HJD} & \multicolumn{1}{c}{$J$} & \multicolumn{1}{c}{$\sigma_J$} & \multicolumn{1}{c}{$H$} & \multicolumn{1}{c}{$\sigma_H$} & \multicolumn{1}{c}{$K_S$} & \multicolumn{1}{c}{$\sigma_K$} \\
\multicolumn{1}{c}{-2450000} &&&&&& \\
\hline
cep001\\
2353.49644 & 15.781 & 0.009 & 15.271 & 0.010 & 15.056 & 0.017 \\
2436.50394 & 15.352 & 0.008 & 14.866 & 0.008 & 14.616 & 0.012 \\
2440.50422 & 15.359 & 0.009 & 14.925 & 0.013 & 14.721 & 0.016 \\
2441.50423 & 15.386 & 0.008 & 14.865 & 0.008 & 14.683 & 0.021 \\
2441.50423 & 15.382 & 0.008 & 14.883 & 0.008 & 14.705 & 0.013 \\
2507.29760 & 15.883 & 0.009 & 15.439 & 0.012 & 15.225 & 0.014 \\
2808.46756 & 15.310 & 0.009 & 14.871 & 0.010 & 14.652 & 0.017 \\
2809.38435 & 15.383 & 0.006 & 14.899 & 0.008 & 14.685 & 0.008 \\
2529.28608 & 15.509 & 0.008 & 15.075 & 0.009 & 14.903 & 0.010 \\
2882.34804 & 15.759 & 0.008 & 15.356 & 0.009 & 15.167 & 0.014 \\
3093.62506 & 15.844 & 0.008 & 15.335 & 0.009 & 15.125 & 0.016 \\
3173.44424 & 15.410 & 0.006 & 14.917 & 0.007 & 14.706 & 0.008 \\
3243.35845 & 15.785 & 0.007 & 15.349 & 0.007 & 15.178 & 0.008 \\
3259.26118 & 15.583 & 0.007 & 15.171 & 0.008 & 15.004 & 0.009 \\
3260.26279 & 15.578 & 0.006 & 15.165 & 0.008 & 15.002 & 0.009 \\
3293.28920 & 15.451 & 0.009 & 14.963 & 0.008 & 14.750 & 0.010 \\
3531.55302 & 15.419 & 0.006 & 14.951 & 0.006 & 14.747 & 0.008 \\
3533.40996 & 15.445 & 0.006 & 14.960 & 0.007 & 14.727 & 0.008 \\
3612.30425 & 15.884 & 0.007 & 15.446 & 0.008 & 15.256 & 0.009 \\
\hline

\end{tabular}
\label{tab_obs}
\end{center}
Table included in full in electronic version.
\end{table}

\begin{table*}
\caption[]{Flux-weighted mean $JHK_S$ for Cepheids in NGC 6822.}
\begin{center}
\begin{tabular}{cccccccccrcc}
\hline
 $J$  &  $\sigma_J$ & $H$ & $\sigma_H$ & $K$& $\sigma_K$ &  $J-H$&$H-K$&$J-K$ &  P (days)  &  name$^a$  & IRSF \\
\hline
15.585 & 0.015 & 15.117 & 0.013 & 14.919 & 0.013 & 0.468 & 0.198 & 0.666 & 123.9000 & cep001  & 10170 \\
15.620 & 0.009 & 15.161 & 0.008 & 15.004 & 0.010 & 0.459 & 0.157 & 0.616  & 65.3200 & cep002  & 30131 \\
16.536 & 0.009 & 16.041 & 0.008 & 15.881 & 0.013 & 0.495 & 0.160 & 0.655  & 37.4610 & cep003  & 10463 \\
16.732 & 0.012 & 16.236 & 0.008 & 16.104 & 0.014 & 0.496 & 0.132 & 0.628  & 34.6630 & cep004  & 40353 \\
16.799 & 0.013 & 16.317 & 0.012 & 16.132 & 0.013 & 0.482 & 0.186 & 0.667  & 30.5120 & cep007  & 11214 \\
17.484 & 0.017 & 17.030 & 0.010 & 16.876 & 0.017 & 0.454 & 0.155 & 0.608  & 19.9600 & cep010  & 30518 \\
17.597 & 0.010 & 17.012 & 0.009 & 16.902 & 0.029 & 0.585 & 0.110 & 0.695  & 19.8870 & cep011  & 12406 \\
17.679 & 0.017 & 17.141 & 0.013 & 17.037 & 0.027 & 0.538 & 0.104 & 0.641  & 19.6020 & cep012  & 30994 \\
17.566 & 0.008 & 17.139 & 0.008 & 16.948 & 0.017 & 0.427 & 0.191 & 0.618  & 18.3390 & cep014  & 40553 \\
17.662 & 0.012 & 17.239 & 0.013 & 17.114 & 0.024 & 0.424 & 0.124 & 0.548  & 17.3440 & cep015  & 11791 \\
17.886 & 0.028 & 17.373 & 0.031 & 17.254 & 0.040 & 0.514 & 0.118 & 0.632  & 16.9600 & cep016  & 30954 \\
17.918 & 0.020 & 17.327 & 0.015 & 17.153 & 0.025 & 0.590 & 0.174 & 0.765  & 16.8550 & cep017  & 12137 \\
17.958 & 0.012 & 17.505 & 0.017 & 17.295 & 0.026 & 0.453 & 0.210 & 0.663  & 13.8720 & cep018  & 40491 \\
18.522 & 0.041 & 18.057 & 0.050 & 17.974 & 0.074 & 0.465 & 0.084 & 0.548  & 10.2770 & cep022  & 31807 \\
18.490 & 0.027 & 18.041 & 0.034 & 17.953 & 0.058 & 0.449 & 0.088 & 0.537  & 9.3664 & cep024 & 13520 \\
18.140 & 0.026 & 17.406 & 0.029 & 17.150 & 0.024 & 0.734 & 0.256 & 0.990  & 8.9367 & cep025 & 12507 \\
18.528 & 0.029 & 18.071 & 0.031 & 17.906 & 0.060 & 0.457 & 0.165 & 0.622  & 8.4670 & cep026 & 31757 \\
17.518 & 0.013 & 16.896 & 0.010 & 16.704 & 0.020 & 0.622 & 0.192 & 0.814  & 7.2085 & cep028 & 31048 \\
18.719 & 0.026 & 17.971 & 0.030 & 17.702 & 0.057 & 0.748 & 0.268 & 1.017  & 2.5937 & cep101 & 31703 \\
\hline
\multicolumn{12}{l}{$^a$Stars cep003 to cep018 used in solutions.}\\ 
\end{tabular}

\end{center}
\label{tab_mean}
\end{table*}

In fitting a line to these data we have omitted  cep001 (period = 123.9 days). Cepheids
of such long periods are known to deviate from extrapolations of linear PL relations
and these stars have been omitted by other observers on these grounds.
In addition we have omitted cep002 (log P = 1.82 days) since it is of longer period
than the LMC stars in the OGLE survey we use (see below). We have also 
omitted the six stars with $\log P < 1.1$. Of these stars cep026 was rejected by Pietrz\'{n}ski
et al (2004) because it deviated from their optical PL relations. Their optical data
also show that cep025, cep028  and cep101 are far too bright for their PL relations 
and they do not include them in their figs. 9, 10 and 11. Our
$JHK_{S}$ data are similarly too bright for our PL relations. Some or all of these stars may
be overtone pulsators. Finally, cep022 and cep024 were omitted because the uncertainty of our
$K_{S}$ is large for these stars. All the data of Table \ref{tab_mean} are shown in the
$K_{S} - \log P$ plot of Fig. \ref{fig_PL}. For illustration the line in this figure
is fitted to the Cepheids chosen above with a slope from the LMC (equation 7 below).  

\section{Discussion}
Previous workers on the Cepheids in NGC 6822 have adopted PL
relations derived for LMC Cepheids and applied these to the
NGC 6822 Cepheids to find the difference between the distance moduli
for the two galaxies. An absolute distance modulus then
follows from an adopted LMC modulus of 18.50 mag. We follow the same general 
procedure here. As done previously by others we obtain an estimate
of the true modulus of NGC 6822 from  the relation between the
apparent moduli at various wavelengths with the relative absorption
coefficients at these wavelengths. A true modulus can also
be obtained from the reddening free parameter $W_{VI}=I-1.55(V-I)$ (see e.g. Udalski et al. 2000).
This relation has the advantage that it also corrects, at least to
 first order, for the intrinsic spread in magnitude and colour at
a given period (width of the instability strip). 
This width is particularly significant at optical wavelengths.
We also use the 
infrared reddening-free parameter $W_{JK_{S}} = K_{S} -0.68 (J-K_{S})$ (see e.g. Persson et al. 2004). 
A brief discussion of the value of the LMC modulus and the question
of metallicity dependence of PL relations is given later.

\subsection{LMC Period-Luminosity Relations}
The most extensive study of LMC Cepheids in $VI$ has
been the work of the OGLE group (e.g.  Soszy\'{n}ski et al. 2008). 
Pietrzy\'{n}ski et al. (2004) adopted OGLE relations in
PL(V), PL(I) and PL($W_{VI})$, specifically those tabulated by 
Udalski (2000) (which are given in a form corrected for adopted
reddenings). These results have been adopted by Gieren et al. (2006)
and Madore et al. (2009a). The latter two papers also adopt LMC PL
relations in $JHK$ from Persson et al. (2004). Madore et al. (2009a)
also adopt LMC PL relations in $[3.6],[4.5],[5.8], [8.0] \mu m$
from Madore et al. (2009b).  

These procedures have the disadvantage
that a joint analysis of NGC 6822 at different wavelengths relies
on LMC relations based on different samples of LMC Cepheids in the
optical, near-IR and mid-IR. This  is particularly so in that the near- and
mid-IR Cepheid sets contain stars lying outside the area of the OGLE
survey, including Cepheids in the NE of the LMC, which is known to be nearer to
us than the main body. It also has the disadvantage that 
it cannot be guaranteed that the mean reddening is the same for
the different samples.

 For the present work we have derived LMC PL relations 
(uncorrected for reddening) for a set of Cepheids common to the OGLEIII
(Soszy\'{n}ski et al. 2008) $VI$, the Persson et al. (2004) near-IR and the Madore et al. (2009b) mid-infrared
samples. It should be noted that whilst the OGLE and Persson et al. data
are intensity mean magnitudes based on full light curves,
the LMC data of Madore et al. are based on observations at two epochs only.
Since the $JHK_{S}$ observations of NGC 6822 Cepheids which we will use to
derive a modulus are all of periods longer than 10 days we also restrict our
LMC sample to the range $1.0< \log P <1.7$. The upper limit is imposed because longer
period Cepheids can deviate from PL relations.
These relations are:
\begin{equation}
V = -2.857 (\pm 0.395) (\log P -1.2) + 14.259 (\pm 0.045)
\end{equation}
\begin{equation}
I = -3.062 (\pm 0.284) (\log P -1.2) + 13.338 (\pm 0.032)
\end{equation}
\begin{equation}
W_{VI} = -3.379 (\pm  0.274) (\log P -1.2) + 11.910 (\pm 0.031)
\end{equation}
\begin{equation}
J = -3.160 (\pm 0.202)(\log P -1.2) + 12.703(\pm 0.023)
\end{equation}
\begin{equation}
H = -3.218(\pm 0.168) (\log P -1.2) + 12.308 (\pm 0.019)
\end{equation}
\begin{equation}
K_{S} = -3.234 (\pm 0.155) (\log P -1.2) + 12.202 (\pm 0.017)
\end{equation}
\begin{equation}
W_{JK_{S}}  = 3.285(\pm 0.132) (\log P -1.2) + 11.862 (\pm 0.015)
\end{equation}
\begin{equation}
[3.6] = -3.244(\pm 0.179) (\log P -1.2) + 12.087 (\pm 0.020)
\end{equation}
\begin{equation}
[4.5] = -3.162 (\pm 0.183) (\log P -1.2) + 12.099 (\pm 0.021)
\end{equation}
\begin{equation}
[5.8] = -3.308 (\pm 0.182) (\log P -1.2) + 12.060 (\pm 0.021)
\end{equation}
\begin{equation}
[8.0] = -3.308 (\pm 0.189) (\log P -1.2) + 12.037 (\pm 0.021)
\end{equation}
The joint sample used to derive these equations contained 32 Cepheids.
Table \ref{tab_LMCcep} in the Appendix lists the stars and their coordinates for future reference.
A joint sample involving shorter period Cepheids and hence more stars could no doubt be constructed. 
However, by restricting  the LMC sample to longer periods we avoid any problems connected with nonlinearity
of the relations. 
The difference between these results and earlier
work is best seen by comparing $W_{VI}$ and $W_{JK_{S}}$ relations which are free
of the problems related to reddening corrections employed by earlier workers.
In the case of the Persson et al $W_{JK_{S}}$ this difference  (at a mean $\log P =1.4$,
which is close to the mean period of the NGC 6822 sample) is 0.073 mag, our result being fainter.
Much of this difference must be due to confining the sample to stars with OGLE data. This is shown
by the fact that the difference between equation 3 for $W_{VI}$ and that used by Pietrzy\'{n}ski et al.
(based on an OGLE LMC relation from Udalski (2000)) is only 0.014 mag (our result being brighter)
at $\log P = 1.4$. Similarly our result (equation 3) is 0.019 mag brighter at this $\log P$
than the latest OGLE III relation (Soszy\'{n}ski et al. (2008)).

\subsection{NGC 6822 $JHK_{S}$ Cepheid data}
\subsubsection{Comparison with previous work}
\begin{figure}
\includegraphics[width=8.5cm]{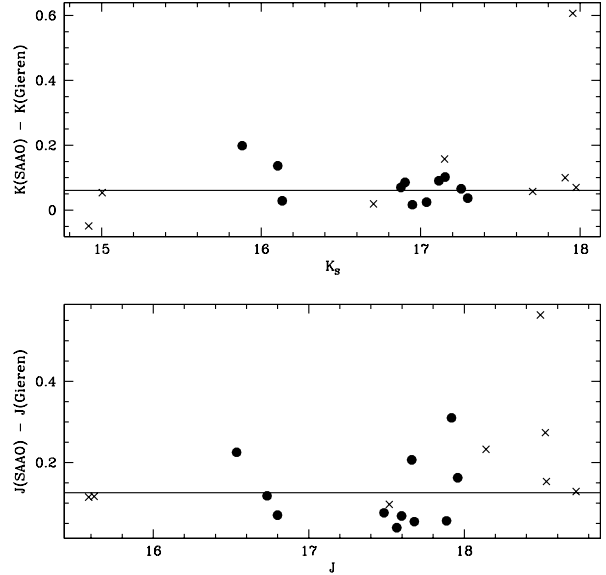}
\caption{Comparison of present mean magnitudes with those of Gieren et al. (2006), corrected to the 2MASS system, plotted against our mean magnitudes. The lines show the mean differences of 0.126 mag in $J$ and 0.060 mag in $K_S$, respectively, for the Cepheids. Those used in our period-luminosity fitting are marked as black dots, the remainder as crosses.
}
\label{fig_diff}
\end{figure}

$JK$ observations of NGC 6822 Cepheids have been made by Gieren et al. (2006).
These are in the UKIRT system. Since these authors had only a small number of observations
per star they derive intensity mean magnitudes by a phase-correction method
based on the optical data of Pietrzy\'{n}ski et al. (2004). These results need to be converted
to the 2MASS system to be compared with our work. For this we have used the relations derived by
Carpenter (2001 as updated on the 2MASS web page). The transformations are small, being
$K_{S}^{2M} -K^{UK} = +0.005$ mag and $J^{2M} - J^{UK} = +0.030$ mag at the mean colour
of the Cepheids compared. We use stars cep003 to cep018 of Table \ref{tab_mean} for the comparison. This omits stars
which lie off the PL relation or have large photometric errors. For the chosen sample the differences,
in the sense IRSF--Gieren, are $\Delta K_{S} = +0.061 \pm 0.014$ mag and $\Delta J = + 0.126 \pm 0.022$ mag.
The results for individual stars are plotted in Fig. \ref{fig_diff}.
These differences evidently need further investigation. In view of them we restrict ourselves in 
the following to an analysis of our own $JHK_{S}$ data so far as the near infrared is concerned,
since we have found no evidence for scale errors in our analysis.

\subsubsection{Conversion to Persson system}
To compare our NGC 6822 results with the Persson results for the LMC discussed in section 4.1, 
they have to be converted to the LCO (NICMOS) system. This was also done using the Carpenter
transformations. At the mean colours of the NGC 6822 Cepheids we use for distance estimation,
the corrections, LCO -- IRSF(2MASS), are small, +0.013 at $J$, +0.008 mag at $H$ and +0.014 mag at $K_{S}$.

In calculating the difference in modulus between NGC 6822 and the LMC we have used the 11 stars,
cep003 to cep018, in Table \ref{tab_mean} as  indicated previously. Including cep002 (P = 65.3 days)
makes no significant difference to our conclusions.  

\subsection{NGC 6822 modulus from $VIJHK_{S}$}
\begin{figure}
\includegraphics[width=8.5cm,viewport=0 0 600 300,clip]{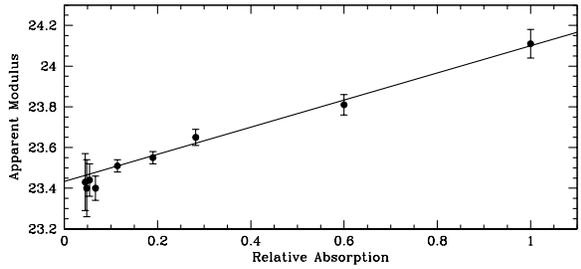}
\caption{Apparent $VIJHK_{S}$ and mid-IR distance moduli determined from NGC 6822 Cepheids plotted as a function of relative absorption as listed in Table \ref{tab_mod}. 
}
\label{fig_mod}
\end{figure}

Assuming the slopes and zero-points given in equations 1 to 11
and a true mean distance modulus of 18.50 for the LMC Cepheid sample we derive the apparent distance moduli
of NGC 6822 listed in Table \ref{tab_mod}. This table also contains relative absorption coefficients based on the Cardelli
 et al. (1989) reddening law and as give by Indebetouw et al. (2005) for the mid-IR.
Fig. \ref{fig_mod} shows a plot of the $VIJHK_{S}$  as well as the mid-infrared apparent moduli against the relative absorption.\\

A least squares solution of the $VIJHK_{S}$ data with equal weight to each point yields a true modulus of $23.43 \pm 0.02$ (int.) mag
and $A_{V} = 0.667$ mag for the amount that the mean visual extinction for the NGC 6822 stars is
greater than that for the fiducial LMC sample. This modulus is entered in Table \ref{tab_truemod}
together with the true moduli derived from $W_{VI}$ and $W_{JK_{S}}$. The errors given combine those
of the LMC relations used (not including any error in the LMC distance modulus)  with those of the NGC 6822 results except in the case of the
least squares fit to the results at the five wavelengths where the internal error is quoted. This latter result shows how closely
the results are fitted by a linear relation. 

An unweighed mean of the three values 
of the true modulus is 23.40 mag, which we take
as our best estimate. It is difficult to estimate the true uncertainty of this value. Even with full light
curves there must be some correlation between the deviations from mean PL relations for a given Cepheid
at different wavelengths and this may have an effect when the number of Cepheids in a sample is
relatively modest. The problem is, of course, worse when one relies on single measurements of a star taken 
simultaneously at different wavelengths (as in the mid-IR work). We estimate an uncertainty of $\pm 0.05$ mag,
not taking into account any error in the adopted distance modulus of the LMC (see below).

\begin{table}
\caption{Cepheid Apparent Moduli of NGC 6822}
\begin{center}
\begin{tabular}{cccccc}
\hline
Band & App Mod & s.e.$^a$ & s.e.$_T^b$ & N$^c$ & $A_{\lambda}/3.1^d$\\
\hline
$V$ & 24.11 & 0.06 & 0.07 & & 1.000\\
$I$ & 23.81 & 0.04 & 0.05 &  & 0.600\\
$J$ & 23.65 & 0.03 & 0.04 & & 0.282\\
$H$ & 23.55 & 0.02 & 0.03 &  & 0.190\\
$K_{S}$ & 23.51 & 0.03 & 0.03 & & 0.114\\
$[3.6]$ & 23.40 & 0.06 & 0.06 & 14 &  0.067\\
$[4.5]$ & 23.44 & 0.07 & 0.08 & 12 &  0.054\\
$[5.8]$ & 23.40 & 0.14 & 0.14 & 6 & 0.048\\
$[8.0]$ & 23.43 & (0.14) & (0.14) & 2 & 0.045\\
\hline
\multicolumn{6}{l}{$^a$ s.e. is the standard error of the NGC 6822 result.}\\
\multicolumn{6}{l}{$^b$ s.e.$_T$ includes the uncertainty in the LMC relations (not}\\
\multicolumn{6}{l}{considering any error in the adopted LMC modulus).} \\
\multicolumn{6}{l}{$^c$ For the mid-IR data the number of stars used.}\\
\multicolumn{6}{l}{$^d$ The adopted relative extinctions.}\\
\end{tabular}
\label{tab_mod}

\end{center}
\end{table}

\begin{table}
\caption{True Modulus of NGC 6822}
\begin{center}
\begin{tabular}{ccc}
\hline
Band & Mod & Note\\
\hline
$VIJHK_{S}$ & $23.43\pm 0.02$ & 1\\
$W_{VI}$ & $23.34 \pm 0.04$ & 2\\
$W_{JK_{S}}$ & $23.42 \pm 0.03$ & 2\\
\hline
Mean & 23.40 & \\
\hline
$[3.6]$ & $23.36 \pm 0.06$ & 2,3 \\
$[4.5]$ & $23.40 \pm 0.08$ & 2,3 \\
$[5.8]$ & $23.37 \pm 0.14$ & 2,3 \\
$[8.0]$ & $23.40 \pm (0.14)$ & 2,3 \\
\hline   
Mean (mid-IR) & 23.38 & \\
\hline
\multicolumn{3}{l}{1. s.e. is internal value}\\
\multicolumn{3}{l}{2. s.e. includes uncertainty in LMC relation}\\
\multicolumn{3}{l}{3. assumes $A_{V} = 0.667$}\\
\end{tabular}
\label{tab_truemod}

\end{center}
\end{table}

\subsection{The NGC 6822 Mid-IR data}
Table \ref{tab_mod} also contains the apparent moduli derived from the mid-IR data of Madore et al (2009a) in the same
way as that described in the previous section. Here we have used all the Cepheids in Madore et al. (2009b)
with $\log P < 1.7$.
We did not use these data in the least squares solution of the last subsection
for a number of reasons: (1) There are no mid-IR data for some of the Cepheids in our 11-Cepheid sample.
It would have reduced the available data too much to confine the solutions to stars in common in the optical,
near-IR and mid-IR samples. For instance, for $[3.6]$ there are only 7 stars in common with our adopted 11. 
(2) The mid-IR data are for single phase measures only. 
The residuals at the various wavelengths are likely to be highly correlated (unless dominated by observational error).
(3) a minor point is that
the reddening law in the mid-IR is not very certain at present. 
The true moduli derived from these data are given in Table \ref{tab_truemod} assuming the difference
in $A_{V}$ between NGC 6822 and the LMC to be 0.667 mag as derived above. 
An unweighted mean of the four values is given in Table \ref{tab_truemod}. It agrees remarkably well
(better than could have been expected) with the results from shorter wavelengths.  
\subsection{Distance Modulus of NGC 6822}
As already noted we adopt 24.40 mag as our best estimate of the NGC 6822 modulus.
The results in Table \ref{tab_truemod} show that with our new $JHK_{S}$ data and revised discussion
the range of derived moduli has been reduced by a factor of two to 0.09 mag. It is interesting to note
that our $W_{VI}$ result agrees exactly with that of Pietrzy\'{n}ski et al. (2004), who included shorter period 
Cepheids in their NGC 6822 sample. As noted above their LMC relation is close to the one we use. The
$W_{VI}$ result is rather sensitive to any systematic errors in the photometry (a zero-point error of
0.02 mag in $I$ leads to a systematic error of 0.05 mag in $W_{VI}$). Errors of this amount are
indeed possible as discussed by Pietrzy\'{n}ski et al. In view of this we do not
consider the difference between the $W_{VI}$ and $W_{JK_{S}}$ results of Table \ref{tab_truemod} to 
be significant.\\
\subsection{A Note on the Adopted LMC Modulus and Metallicity Effects}
The derived distance modulus for NGC 6822 assumes an LMC modulus of 18.50 mag. Various distance indicators give
values close to this (see for instance recent summaries by Feast (2012), and Walker (2011)). With a zero-point based on a reduced parallax type solution
of the parallaxes of Galactic Cepheids (Benedict et al. 2007; van Leeuwen et al. 2007 ), an  LMC modulus of $18.52 \pm 0.03$ mag was obtained for $W_{VI}$ from an OGLE Cepheid sample. Similarly, a modulus of $18.47 \pm 0.03$ mag was obtained
using the Persson et al. (2004) $K_{S}$ results for the LMC. The difference between these two estimates
is not significant. However, it is interesting to note that a difference in this sense is expected
in view of the discussion of section 4.1.
In neither case was any correction made for metallicity
differences between the LMC and Galactic samples. The metallicity of a young (i.e. Cepheid-like) population
in NGC 6822 is intermediate between that of the LMC and the SMC (see Venn et al. (2001) for a discussion).
No metallicity corrections have been applied to our derived NGC 6822 moduli. Whether there are
significant metallicity effects on Cepheid relations at different wavelengths remains somewhat
controversial.
It should also be noted that whilst the relative distances
of the LMC and NGC 6822 rest on a comparison of long period Cepheids, the LMC Cepheid distance 
comes primarily from a comparison of short period Cepheids with Galactic Cepheids of known parallax.
\subsection{The RR Lyrae Distance Modulus of NGC 6822}
An RR Lyrae distance modulus to NGC 6822 can be derived from the work of Baldacci et al. (2004) and Clementini
et al. (2003). The 24 ab-type RR Lyraes in table 4 of Baldacci et al. have mean magnitude, $V =24.63\pm 0.04$ mag. From
a period-metallicity relation Clementini et al. estimate that the NGC 6822 RR Lyraes have a mean $[Fe/H] = -1.92$.
At this metallicity the calibration of an $M_{V} - [Fe/H]$ relation from the parallaxes of
Galactic RR Lyrae variables (Benedict et al. 2011) gives $M_{V} = 0.37 \pm 0.04$ mag and an apparent
distance modulus of $(m-M) = 24.26$ mag. If we adopt $A_{V} = 0.77$ mag from Clementini et al., based on
Schlegel et al. (1998), then $(m-M)_{0} =23.49$ mag. (This differs from the Clementini et al. result, 23.36, 
almost entirely because their RR Lyrae zero point is fainter than the Benedict et al. one used here.)
The estimate of the absorption is for the foreground of
NGC 6822 only and uncertainty in the absorption may be the main uncertainty in this
result. Without taking this uncertainty or that in the metallicity estimate into account 
the standard error of the modulus is $\pm 0.06$ mag. The agreement with the Cepheids uncorrected
for metallicity effects is evidently satisfactory. 
\section{Conclusions}
NGC 6822 is a test case for the precision which can be achieved in practice in deriving
extragalactic distances from classical Cepheids. 
As in most current extragalactic work the Cepheid distance to NGC 6822 is derived relative to
the LMC Cepheids.
The LMC is known to have significant depth and structure in the line of sight
and reddenings vary from star to star. Complications
can then arise, especially in combining results at different wavelengths. Fiducial
period-luminosity relations were therefore derived for a common set of long period
Cepheids at wavelengths from the optical to the mid-infrared. Appreciable differences
exist between these relations and some used earlier.
Using the new relations together with new multi-epoch observations of NGC 6822 Cepheids in 
$JHK_{S}$ reduces the spread of moduli derived in different ways by a factor of two
compared to earlier work.
Our best estimate of the true modulus of NGC 6822 is 23.40 mag. The standard error of this
quantity is estimated as $\sim0.05$ mag (i.e. a 3\% distance scale uncertainty)
not taking into account any metallicity effects on either the LMC or the NGC 6822 Cepheids.

A distance modulus of the galaxy from RR Lyrae variables and based on a recent
calibration of the RR Lyrae scale is 23.49 mag, which agrees with the Cepheid result
within the uncertainties.

Work is in progress on the Mira variables in NGC 6822
and on the infrared colour-magnitude diagram. This will allow further comparison of distance estimates. 

\section*{Acknowledgements}
This publication makes use of data products from the Two Micron All Sky Survey, which is a joint project of the University of Massachusetts and the Infrared Processing and Analysis Center/California Institute of Technology, funded by the National Aeronautics and Space Administration and the National Science Foundation.

MWF and PAW gratefully acknowledge the receipt of research grants from the National Research Foundation (NRF) of South Africa.

\appendix
 \section{LMC Cepheids for PL Relations}      
\clearpage

\begin{table}
\caption{Cepheids used for LMC PL relations.}
\begin{tabular}{llll}
\hline
OGLE Name & HV Name & \multicolumn{2}{c}{$\alpha\ \ \  (2000.0)\ \ \ \delta$ }\\
\hline
OGLE-LMC-CEP-0070 & HV12724 & 04:46:01.08 & -69:38:55.8\\
OGLE-LMC-CEP-0174 & HV12471 & 04:50:52.43 & -69:18:55.9\\
OGLE-LMC-CEP-0467 & HV876   & 04:57:12.34 & -67:22:57.3\\
OGLE-LMC-CEP-0500 & HV2244  & 04:57:50.87 & -67:50:18.9\\
OGLE-LMC-CEP-0501 & HV878   & 04:57:51.03 & -69:57:29.7\\
OGLE-LMC-CEP-0504 & HV12505 & 04:57:56.73 & -68:48:57.6\\
OGLE-LMC-CEP-0648 & HV2270  & 05:00:48.36 & -69:31:54.7\\
OGLE-LMC-CEP-0655 & HV2260  & 05:00:55.86 & -68:26:20.8\\
OGLE-LMC-CEP-0683 & HV2282  & 05:01:24.94 & -70:04:18.3\\
OGLE-LMC-CEP-0727 & HV887   & 05:02:10.24 & -69:32:23.7\\
OGLE-LMC-CEP-0819 & HV2291  & 05:03:46.16 & -68:52:36.4\\
OGLE-LMC-CEP-0821 & HV889   & 05:03:49.50 & -68:56:02.7\\
OGLE-LMC-CEP-0844 & HV891   & 05:04:15.47 & -69:01:36.4\\
OGLE-LMC-CEP-0848 & HV892   & 05:04:21.08 & -68:43:42.8\\
OGLE-LMC-CEP-0935 & HV893   & 05:06:00.89 & -69:06:17.1\\
OGLE-LMC-CEP-0986 & HV899   & 05:07:07.81 & -68:53:19.5\\
OGLE-LMC-CEP-1001 & HV2324  & 05:07:21.69 & -68:20:18.3\\
OGLE-LMC-CEP-1031 & HV901   & 05:07:42.13 & -69:14:48.1\\
OGLE-LMC-CEP-1058 & HV904   & 05:08:18.27 & -68:46:47.1\\
OGLE-LMC-CEP-1088 & HV2339  & 05:08:49.54 & -68:59:59.1\\
OGLE-LMC-CEP-1184 & HV5655  & 05:11:05.41 & -70:30:34.4\\
OGLE-LMC-CEP-1538 & HV2432  & 05:18:13.79 & -68:19:30.5\\
OGLE-LMC-CEP-1578 & HV932   & 05:19:14.80 & -69:36:18.1\\
OGLE-LMC-CEP-1954 & HV2527  & 05:25:39.09 & -71:06:39.9\\
OGLE-LMC-CEP-2023 & HV2549  & 05:27:00.58 & -71:38:35.8\\
OGLE-LMC-CEP-2030 & HV2538  & 05:27:07.76 & -68:29:42.9\\
OGLE-LMC-CEP-2337 & HV997   & 05:33:00.98 & -68:11:27.6\\
OGLE-LMC-CEP-2534 & HV1005  & 05:36:06.80 & -68:49:13.4\\
OGLE-LMC-CEP-2636 & HV1006  & 05:37:22.45 & -69:28:59.4\\
OGLE-LMC-CEP-2949 & HV2793  & 05:41:48.53 & -68:41:16.1\\
OGLE-LMC-CEP-3013 & HV1019  & 05:42:51.05 & -70:08:12.3\\
OGLE-LMC-CEP-3203 & HV12656 & 05:48:06.94 & -71:30:21.4\\
\hline
\end{tabular}
\label{tab_LMCcep}
\end{table}


\begin{thebibliography}{}

\bibitem[]{baldacci} Baldacci L., Rizzi L., Clementini G., Held E.V., 2005, A\&A, 431, 1189
\bibitem[]{benedict07} Benedict G.F et al., 2007, AJ, 133, 1810
\bibitem[]{benedict11} Benedict G.F. et al., 2011, AJ, 142, 187
\bibitem[]{cardelli} Cardelli J.A., Clayton G.C., Mathis J.S., 1989, AJ, 45, 245
\bibitem[]{carpenter} Carpenter J.M., 2001, AJ, 121, 2851 
\bibitem[]{clementini} Clementini G., Held E.V., Baldacci L., Rizzi L., 2003, ApJ, 588, L85
\bibitem[]{feast} Feast M.W., 2012, in Planets, Stars and Stellar Systems, vol 5,
Stellar Systems and Galactic Structure ed. G. Gilmore, Springer, Berlin, in press
\bibitem[]{gieren} Gieren W. et al., 2006, ApJ, 647, 1056
\bibitem[]{indebetouw} Indebetouw R. et al., 2005, ApJ, 619, 931
\bibitem[]{madore09a} Madore B.F., Rigby J., Freedman W.L., Persson S.E., Sturch L., Mager V., 2009a, ApJ, 693, 936
\bibitem[]{madore09b} Madore B.F., Freedman W.L., Rigby J., Persson S.E., Sturch L., Mager V., 2009b, ApJ, 695, 988
\bibitem[]{persson} Persson S.E., Madore B.F., Krzemi\'{n}ski W., Freedman W.L., Roth M., Murphy D.C., 2004, AJ, 128, 2239
\bibitem[]{pietrzynski} Pietrzy\'{n}ski G. et al. 2004, AJ, 128, 2815
\bibitem[]{riess} Riess A.G., Fliri J., Valls-Gabaud D., 2011, arXiv:1110.3769
\bibitem[]{schecter} Schechter P. L., Mateo M., Saha A., 1993, PASP, 105, 1342
\bibitem[]{schlegel} Schlegel D.J., Finkbeiner D.P, Davis M., 1998, ApJ, 500, 525 
\bibitem[]{soszynski} Soszy\'{n}ski I. et al., 2008, AcA, 58, 163
\bibitem[]{udalski} Udalski A., 2000, AcA, 50, 279
\bibitem[]{vleeuwin} van Leeuwen F., Feast M.W., Whitelock P.A., Laney C.D., 2007, MNRAS, 379, 723
\bibitem[]{venn} Venn K.A. et al., 2001, ApJ, 547, 765
\bibitem[]{walker} Walker, A.R., 2011, arXiv:1112.3171 



\end{thebibliography}
\end{document}